\documentclass[journal=cmatex,manuscript=article]{achemso}
\usepackage[version=3]{mhchem} 
\usepackage[T1]{fontenc}       
\usepackage[version=3]{mhchem} 
\usepackage[T1]{fontenc}       
\usepackage{graphicx}
\usepackage{dcolumn}
\usepackage{bm}
\usepackage{xcolor}
\usepackage{amsmath}

\newcommand{\CTSO}{Cu$_3$(TeO$_4$)(SO$_4$)$\cdot$H$_2$O}
\newcommand{\C}{$^\circ$C}

\author{Zhi-Cheng~Wang}
\affiliation{Department of Physics, Boston College, Chestnut Hill, MA 02467, USA}
\email{zhicheng.wang@bc.edu}
\author{Kulatheepan~Thanabalasingam}
\affiliation{Department of Chemistry and Biochemistry, University of Texas at Dallas, Richardson, TX 75080, USA}
\author{Jan~P.~Scheifers}
\affiliation{Department of Chemistry and Biochemistry, University of Texas at Dallas, Richardson, TX 75080, USA}
\author{Alenna~Streeter}
\affiliation{Department of Physics, Boston College, Chestnut Hill, MA 02467, USA}
\author{Gregory~T.~McCandless}
\affiliation{Department of Chemistry and Biochemistry, University of Texas at Dallas, Richardson, TX 75080, USA}
\author{Jonathan~Gaudet}
\affiliation{Department of Materials Science and Engineering, Maryland University, College Park, Maryland, 20942-2115, USA}
\alsoaffiliation{NIST Center for Neutron Research, National Institute of Standards and Technology, Gaithersburg, Maryland, 20899-6102, USA}
\author{Craig~M.~Brown}
\affiliation{NIST Center for Neutron Research, National Institute of Standards and Technology, Gaithersburg, Maryland, 20899-6102, USA}
\author{Carlo~U.~Segre}
\affiliation{Department of Physics \& CSRRI, Illinois Institute of Technology, Chicago, IL 60616, USA}
\author{Julia~Y.~Chan}
\affiliation{Department of Chemistry and Biochemistry, University of Texas at Dallas, Richardson, TX 75080, USA}
\author{Fazel~Tafti}
\affiliation{Department of Physics, Boston College, Chestnut Hill, MA 02467, USA}
\email{fazel.tafti@bc.edu}
\phone{(617) 552-2127}
\title{Antiferromagnetic Order and Spin-Canting Transition in the Corrugated Square Net Compound \CTSO}
\abbreviations{AFM, FC, ZFC, XRD , SQUID, EDX}
\keywords{Tellurite, Sulfate, Mott insulator, Hubbard model, Magnetism, Structure, Heat capacity}
\begin{document}







\pagebreak
\begin{abstract}
Strongly correlated electrons in layered perovskite structures have been the birthplace of high-temperature superconductivity, spin liquid, and quantum criticality.
Specifically, the cuprate materials with layered structures made of corner sharing square planar CuO$_4$ units have been intensely studied due to their Mott insulating grounds state which leads to high-temperature superconductivity upon doping.
Identifying new compounds with similar lattice and electronic structures has become a challenge in solid state chemistry.
Here, we report the hydrothermal crystal growth of a new copper tellurite sulfate \CTSO, a promising alternative to layered perovskites. 
The orthorhombic phase (space group $Pnma$) is made of corrugated layers of corner-sharing CuO$_4$ square-planar units that are edge-shared with TeO$_4$ units.
The layers are linked by slabs of corner-sharing CuO$_4$ and SO$_4$.
Using both the bond valence sum analysis and magnetization data, we find purely Cu$^{2+}$ ions within the layers, but a mixed valence of Cu$^{2+}$/Cu${^+}$ between the layers.   
\CTSO\ undergoes an antiferromagnetic transition at $T_N$=67~K marked by a peak in the magnetic susceptibility.
Upon further cooling, a spin-canting transition occurs at $T^{\star}$=12~K evidenced by a kink in the heat capacity.
The spin-canting transition is explained based on a $J_1$-$J_2$ model of magnetic interactions, which is consistent with the slightly different in-plane super-exchange paths. 
We present \CTSO\ as a promising platform for the future doping and strain experiments that could tune the Mott insulating ground state into superconducting or spin liquid states.
\end{abstract}

\pagebreak

\section{\label{introduction}Introduction}
\begin{figure}
 \includegraphics[width=0.55\textwidth]{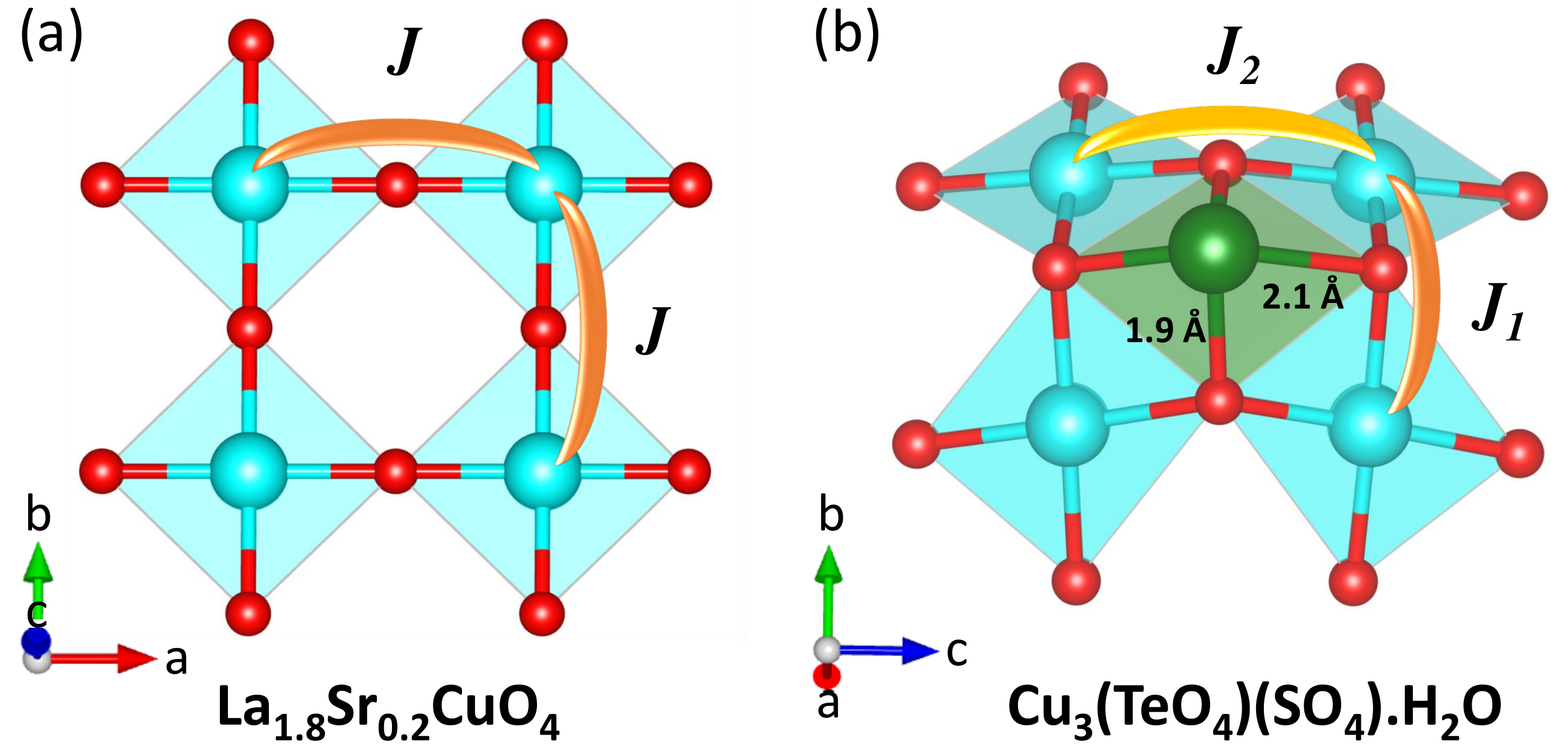}
 \caption{\label{fig:MOTIVATION}
 (a) Corner-sharing square planar CuO$_4$ units with a Cu-O-Cu square net in a representative cuprate superconductor La$_{1.8}$Sr$_{0.2}$CuO$_4$ with a single $J$ coupling.
 (b) The corrugated square net in \CTSO\ with two magnetic coupling constants $J_1$$\geq$$J_2$.}
 \end{figure}

Mott insulators are materials with half-filled bands, a nominally metallic configuration, but with strong correlations that lead to localized electronic states and insulating behavior.
Such materials have been a focus of intense research, largely due to the discovery of high-$T_c$ superconductivity in cuprate systems~\cite{phillips_exact_2020}.
Aside from subtle structural differences, all cuprate superconductors have a layered crystal structure made of square networks of Cu-O-Cu bonds as illustrated in Figure~\ref{fig:MOTIVATION}a.
Each Cu$^{2+}$ with a single hole in the $d_{x^2-y^2}$ orbital acts as a spin-1/2 ion whose strong interaction with the neighboring ions leads to an antiferromagnetic (AFM) insulating ground state~\cite{anderson_twenty-five_2013}.
Since all bond lengths on the square net are equal, a single nearest neighbor magnetic coupling denoted by $J$ on Figure~\ref{fig:MOTIVATION}a can describe the basic magnetic interactions.
This spin model can be mapped onto a charge model known as the Hubbard Hamiltonian, $\mathcal{H}=-t\sum_{<i,j>,\sigma}c_{i,\sigma}^\dagger c_{j,\sigma} + U\sum_i n_{i\uparrow}^\dagger n_{i\downarrow}$, where the first term describes hopping of electrons between neighboring sites with the amplitude $t=4J^2/U$, and the second term accounts for the Coulomb cost of double occupancy on a single site~\cite{anderson_twenty-five_2013,foley_coexistence_2019}.
At half-filling, the Mott insulator orders antiferromagnetically, but upon doping, it undergoes a quantum phase transition into a superconducting state~\cite{lee_doping_2006,jiang_superconductivity_2019}.
Despite several theoretical proposals, finding new Mott insulators with a similar electronic structure to the cuprates has become a major challenge in solid-state chemistry~\cite{baskaran_five-fold_2009}.

Recently, it has been suggested that the copper tellurium oxides and hydroxides~\cite{norman_copper_2018} could be a new playground for magnetism and superconductivity due to the half-filled bands from Cu$^{2+}$ in such compounds as CuTeO$_4$, Sr$_2$CuTeO$_6$, Cu$_3$TeO$_6\cdot$(H$_2$O)$_2$, MgCu$_2$TeO$_6\cdot$(H$_2$O)$_6$, K$_2$Cu$_2$(Te$_2$O$_5$)(TeO$_3$)$_2\cdot$(H$_2$O)$_2$, and Cu$_6$IO$_3$(OH)$_{10}$Cl~\cite{falck_crystal_1978,iwanaga_crystal_1999,grice_jensenite_1996,margison_crystal_1997,mills_bluebellite_2014}.
Although these materials could be magnetic insulators, none of them have a corner-shared square network of CuO$_4$ as shown in Figure~\ref{fig:MOTIVATION}a which is necessary for the Hubbard model.
Furthermore, most of these materials crystallize in the honeycomb, Kagome, and maple-leaf structures~\cite{norman_copper_2018}, except for CuTeO$_4$ and Sr$_2$CuTeO$_6$ that have a cubic unit cell, but without a 2D corner-shared square network~\cite{falck_crystal_1978,iwanaga_crystal_1999}.

In this article, we present a new copper tellurite sulfate material with the chemical formula \CTSO\ whose quasi-2D lattice comprises a distorted square-planar network of Cu-O-Cu bonds as illustrated in Figure~\ref{fig:MOTIVATION}b.
Each spin-1/2 Cu$^{2+}$ ion interacts with its neighbors and creates a Mott insulating state with AFM ordering at $T_N$=67~K, a description that fits the Hubbard model.
Unlike the cuprates, a distinguishable feature of the phase is the buckled square network of Cu-O-Cu bonds due to the presence of a Te$^{4+}$ ion at the center of every four corner-shared CuO$_4$ units (Figure~\ref{fig:MOTIVATION}b).
The two different Te-O bond lengths (1.9 and 2.1~\AA) will translate to two AFM coupling constants denoted by $J_1$ and $J_2$ on Figure~\ref{fig:MOTIVATION}b.
Thus, \CTSO\ is an excellent candidate material for the extended Hubbard model with $J_1$$\geq$$J_2$, due to the slightly shorter distance between Cu$^{2+}$ ions across the $J_1$ super-exhcange path compared to the $J_2$ path~\cite{tocchio_magnetic_2020}.
Our combined magnetization and heat capacity measurements reveal two magnetic transitions due to the two competing coupling constants.
First, \CTSO\ develops an AFM order at $T_N$=67~K, then it undergoes a spin-canting transition at $T^\star$=12~K where the spins deviate from an ideal antiparallel alignment.
Our findings in \CTSO\ revive the search for new compounds with potential for a Mott insulator ground state, extended Hubbard model, and possibly superconductivity.

\section{\label{experimental}Experimental Section}
\subsection{Crystal growth}
\CTSO\ crystals were grown using a hydrothermal method.
The starting materials Cu(NO$_3$)$_2\cdot$2.5H$_2$O (5~g, 21.5~mmol), Na$_2$S$\cdot$9H$_2$O (1.5~g, 6.2~mmol), Te powder (0.5~g, 3.9~mmol), and 4~mL of deionized water were loaded into a 10~mL Teflon liner (90\% full) inside a steel autoclave and mixed with a glass stirring rod.
The autoclave was kept at 220~\C\ for 250 hours in a laboratory oven.
The reaction had a high yield of approximately 1.5~g \CTSO\ crystals which we harvested after washing the solvent and byproducts with deionized water.
The crystals were brittle, had a dark green color, and grew with an acicular habit as shown in the inset of Figure~S1 (Supporting Information). 
The reaction byproducts were Cu$_3$(SO$_4$)(OH)$_4$ (green powder) and Cu$_7$(TeO$_3$)$_2$(SO$_4$)$_2$(OH)$_6$ (turquoise crystals), with a ratio of Cu:Te:S = 3.5:1:1 close to the ratio in \CTSO.

\subsection{X-ray diffraction}
\begin{table}
\caption{\label{tab:tab1}Crystallographic parameters and X-ray refinement statistics summarized for a single crystal of \CTSO.}
\begin{tabular}{ll}
\hline
\hline
Space Group & $Pnma$ (\#62)\\
$a$ (\AA) & 15.974(2)\\
$b$ (\AA) & 6.3468(8)\\
$c$ (\AA) & 7.2563(14)\\
$V$ (\AA$^3$) & 735.7(2)\\
$Z$ & 4\\
Temperature (K)  & 298\\
$\theta$ range (deg) & 2.6-30.6\\
$\mu$ (mm$^{-1}$) & 12.79\\
Measured reflections & 43789\\
Independent reflections & 1220\\
$R_\mathrm{int}$ & 0.100\\
$h$ & $-22\rightarrow 22$\\
$k$ & $-9\rightarrow 9$\\
$l$ & $-10\rightarrow 10$\\
GoF & 1.08\\
Extinction coefficient & 0.00330(17)\\
$R_1(F^2 > 2\sigma(F^2))$~$^a$ & 0.014\\
$wR_2(F^2)$~$^b$ & 0.034 \\
$\Delta\rho_\mathrm{max}$ (e \AA$^{-3}$) & 0.83\\
$\Delta\rho_\mathrm{min}$ (e \AA$^{-3}$) & $-$0.57\\
\hline
\hline
$^a$ $R_1=\sum||F_o|-|F_c||/\Sigma|F_o|$\\
$^b$ $wR_2=\left[ \sum  w \left( F_o^2-F_c^2 \right)^2  / \sum w \left( F_o^2 \right)^2 \right]^\frac{1}{2}$
\end{tabular}
\end{table}
A small single crystal with dimensions $0.03\times0.03\times0.09$ mm$^3$ was selected and data were collected on a Bruker D8 Quest diffractometer equipped with an Incoatec microfocus source (I$\mu$S, Mo $K_\alpha$ radiation, $\lambda$=0.71073 \AA), and a PHOTON II CPAD area detector\footnote{Certain commercial equipment, instruments, or materials are identified in this document. Such identification does not imply recommendation or endorsement by the National Institute of Standards and Technology, nor does it imply that the products identified are necessarily the best available for the purpose.}.
The collected frames were reduced using the Bruker SAINT software, and a multiscan absorption correction was applied using Bruker SADABS~\cite{Bruker}.
The initial structural model was developed with the intrinsic phasing feature of SHELXT~\cite{SHELXT} and a least-square refinement was performed using SHELXL2014~\cite{SHELXT2014}.
%
We observed positional disorder of the solvent water molecules along the $b$-direction, which has been refined as disordered over a split site.
The crystallographic information and refinement statistics are summarized in Table~\ref{tab:tab1}.
The atomic coordinates, displacement parameters are presented in Table~\ref{tab:tab2}.
Selected bond distances and Cu-Cu distances are provided in Table~\ref{tab:tab3}.

\begin{table}
\caption{\label{tab:tab2} The fractional atomic coordinates, site occupancies, and equivalent isotropic displacement parameters ($U_{eq}$) are listed for each Wyckoff site in the structure of \CTSO. The atomic displacement parameters of H1 and H2 were refined isotropically whereas all the other atoms were refined anisotropically.}
\begin{tabular}{lllllll}
\hline
\hline
 Atom & Site &  $x$ & $y$ & $z$ & $U_\mathrm{eq}$~(\AA$^2$) & Occ.      \\  
\hline 
 Te1 & 4$c$    & 0.13721(2) & 1/4 & 0.64673(2) & 0.00792(6) & 1         \\
 Cu1 & 8$d$    & 0.25156(2) & 0.50293(3) & 0.88327(3) & 0.00910(7) & 1  \\
 Cu2 & 4$c$    & 0.40689(2) & 3/4 & 0.67012(5) & 0.01054(8) & 1         \\
 S1 & 4$c$     & 0.57556(4) & 3/4 & 0.41099(9) & 0.00811(13) & 1        \\
 O1 & 4$c$     & 0.32502(12) & 3/4 & 0.8654(2) & 0.0078(4) & 1          \\
 O2 & 4$c$     & 0.18265(12) & 1/4 & 0.9060(2) & 0.0084(4) & 1          \\
 O3 & 8$d$     & 0.20930(10) & 0.4893(2) & 0.63332(17) & 0.0114(3) & 1  \\
 O4 & 8$d$     & 0.59172(10) & 0.5649(2) & 0.2921(2) & 0.0161(3) & 1    \\
 O5 & 4$c$     & 0.48693(12) & 3/4 & 0.4662(3) & 0.0144(4) & 1          \\
 O6 & 4$c$     & 0.63037(13) & 3/4 & 0.5712(3) & 0.0187(4) & 1          \\
 O7 & 4$c$     & 0.5340(2) & 0.717(2) & 0.8939(4)	& 0.032(3) & 0.5    \\
 H1 & 8$d$     & 0.551(3) & 0.675(8) & 0.997(4) & 0.048  & 0.5          \\
 H2 & 8$d$     & 0.579(2) & 0.756(19) & 0.839(6) & 0.048 & 0.5          \\
\hline
\hline
\end{tabular}
\end{table}

\begin{table}
\caption{\label{tab:tab3}Selected bond distances and Cu-Cu distances.}
\begin{tabular}{ll}
\hline
\hline
Te1-O1 & 2.1285(18)  \\          
Te1-O2 & 2.0166(18)   \\         
Te1-O3 ($\times$2) & 1.9087(14)\\
Cu1-O1 & 1.9628(12)   \\         
Cu1-O2 & 1.9535(11)   \\         
Cu1-O3 & 1.9371(13)   \\         
Cu1-O3 & 1.9198(13)   \\         
Cu2-O1 & 1.9286(18)   \\         
Cu2-O4 ($\times$2) & 2.0176(14)\\
Cu2-O5 & 1.955(2)     \\         
S1-O4 ($\times$2) & 1.4801(14)\\ 
S1-O5 & 1.471(2)      \\        
S1-O6 & 1.455(2)      \\        
 \hline
Cu1-Cu1($b$ direction) & 3.1362(6)\\
Cu1-Cu1 ($c$ direction) & 3.6287(8)\\
Cu1-Cu1(diagonal) & 4.8204(6)\\
Cu1-Cu2 (short) & 3.3178(5)\\
Cu1-Cu2 (long) & 3.6491(6)\\
Cu2-Cu2 ($b$ direction) & 6.3468(9)\\
Cu2-Cu2 ($c$ direction) & 7.256(2)\\
Cu2-Cu2 (diagonal) & 5.0015(7)\\
\hline
\hline
\end{tabular}
\end{table}

\subsection{Neutron Diffraction}
Powder neutron diffraction experiment was performed on the BT-1 high resolution powder diffractometer at the NIST Center for Neutron Research. 
Approximately 2 grams of \CTSO\ powder sample was loaded in an aluminum canister, with 1 bar of helium exchange gas loaded at room temperature. 
The powder can was installed into a closed cycle refrigerator with a base temperature of 4.6~K. 
We collected diffraction patterns using 60' collimation and the Ge(311) monochromator ($\lambda$=2.079~\AA) for $T$=4.6, 25, and 100~K. 
All error bars shown in this work indicate one standard deviation.
We could not resolve the magnetic Bragg peaks due to the large background from the incoherent scattering of neutrons by the hydrogen atoms; however, the analysis of neutron data has confirmed the X-ray structural refinement.
Therefore, the neutron data are presented entirely in the Supporting Information.

\subsection{Thermal Analysis}
Thermogravimetric analysis (TGA) was performed using a TA Instruments Discovery SDT650 under a constant flow of N$_2$ (100~mL/min) and a heating rate of 1\C/min to a maximum of 300\C. 
The crushed crystals were placed in an alumina crucible without a lid for the experiment. 
The sample was held at 75\C\ for 30 minutes before ramping to 300\C\ to ensure thermal equilibrium and remove surface moisture.

\subsection{Physical Measurements}
Heat capacity was measured on a crystal cluster of mass 7.9~mg using a Quantum Design PPMS DynaCool with a relaxation-time technique.
The flat surface of the crystal cluster was attached to the sample platform with the Apiezon-N grease.
DC magnetization measurements were performed using a Quantum Design MPMS3 on the same sample that we used for the heat capacity measurements.
%
Energy Dispersive X-ray Spectroscopy (EDX) was performed using an EDAX detector installed on a JEOL-7900F field emission electron microscope (FESEM).
The spectra were obtained from the fresh surface of a crystal and confirmed the chemical formula of the title compound (Supporting Information, Figure~S1).

\section{\label{results}Results and Discussion}

\subsection{\label{synthesis}Synthesis}
Our hydrothermal synthesis of \CTSO\ can be described as a redox reaction where the element Te and anion S$^{2-}$ are oxidized into cations Te$^{4+}$ and S$^{6+}$.
The possibility that O$_2$ as the oxidant is ruled out because the reactants are sealed in the autoclave and the remaining O$_2$ is too little to oxidize Te and Na$_2$S.
Candidate reduction reactions include: 2H$^+\rightarrow$ H$_2$, Cu$^{2+}\rightarrow$ Cu$^+$, and NO$_3^-\rightarrow$ NO$_2^-$.
In the case of 2H$^+\rightarrow$ H$_2$, the reaction would produce an enormous gas pressure ($\sim10^3$ bar) and rupture the blow-off valve on the autoclave.
Since we do not see evidence of such a damage, we rule out this possibility.
If Cu$^{2+}$ is reduced to Cu$^+$, the chemical equation must be 15Cu$^{2+}$(NO$_3$)$_2$ + Na$_2$S + Te + 9H$_2$O $\rightarrow$ Cu$^{2+}_3$(TeO$_4$)(SO$_4$)$\cdot$H$_2$O + 2NaNO$_3$ + 16HNO$_3$ + 12Cu$^+$NO$_3$.
We rule out this possibility too based on three observations;
(i) The product contains Cu$^+$ which is not stable in an oxidizing environment;
(ii) The reaction requires a large amount of Cu(NO$_3$)$_2$ reagent and produces only a small amount of \CTSO, which is inconsistent with the observed high yield of the reaction;
(iii) The reaction completely fails if Cu(NO$_3$)$_2\cdot$2.5H$_2$O is replaced with CuSO$_4\cdot$5H$_2$O as the starting material, which means the reduction of NO$_3^-$ to NO$_2^-$ is crucial in achieving the title compound.
Thus, the redox reaction can only be described as:
\begin{equation}
\label{eq:synthesis}
\mathrm{3Cu(NO_3)_2 + Na_2S + Te + 3H_2O \rightarrow Cu_3(TeO_4)(SO_4)\cdot H_2O + 2NaNO_2 + 4HNO_2}
\end{equation}
which is consistent with both the high-yield of the reaction and the crucial role of Cu(NO$_3$)$_2$ in the synthesis.

\subsection{\label{structural}Structural Analysis}
\begin{figure}
\includegraphics[width=0.55\textwidth]{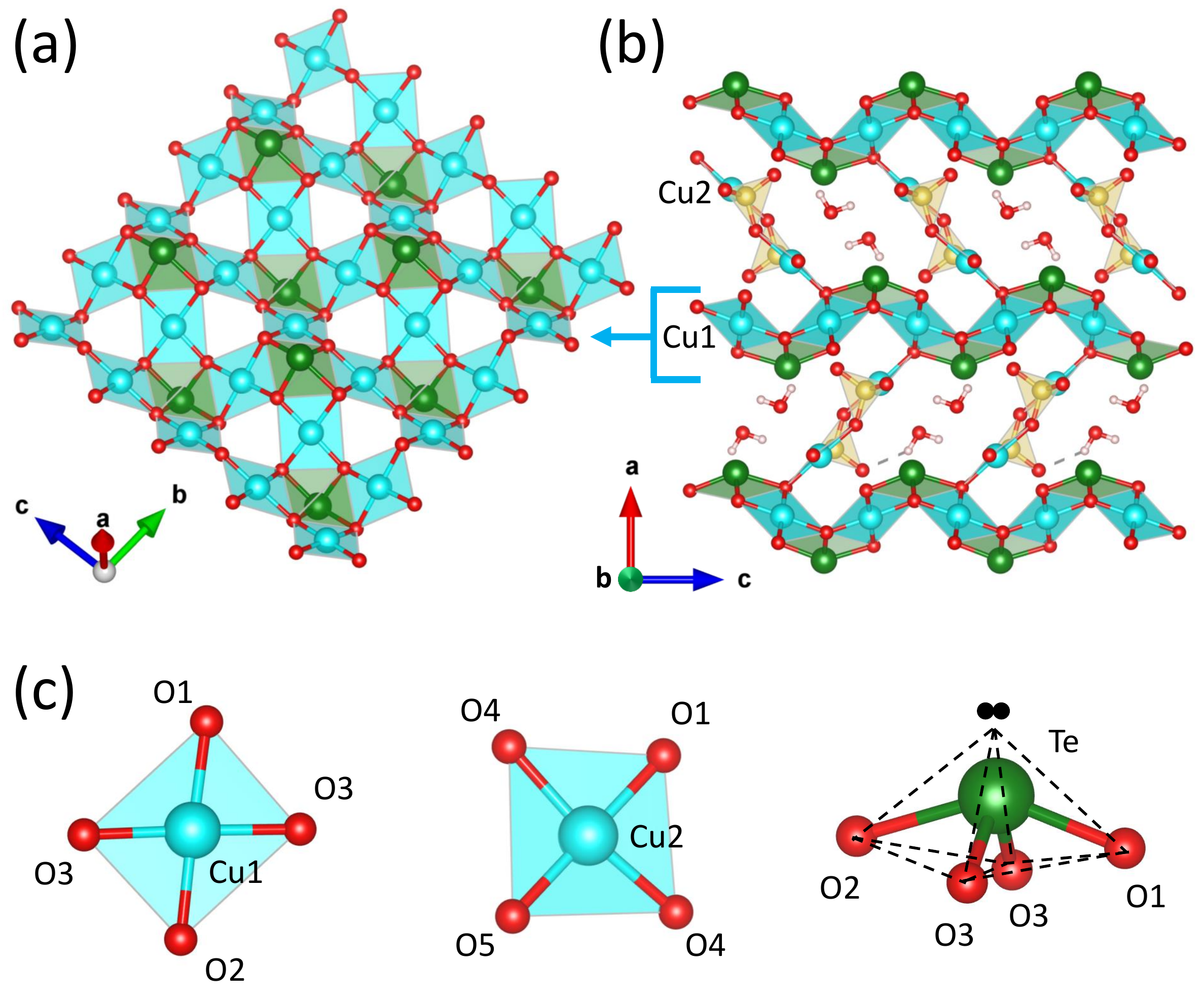}
\caption{\label{fig:CIF}
(a) Corrugated $bc$-planes in the crystal structure of \CTSO\ comprise corner-sharing CuO$_4$ squares and edge-sharing CuO$_4$-TeO$_4$ units.
The Cu, Te, S, and O atoms appear as blue, green, yellow, and red spheres, respectively.
(b) A $b$-axis view shows the corrugated layers linked by corner-sharing CuO$_4$-SO$_4$ slabs with alternating directions.
The H$_2$O molecules reside in the channels parallel to the $b$-axis.
(c) The local distorted square-planar coordination around Cu1 and Cu2 sites, and the trigonal bipyramidal coordination around Te$^{4+}$.
The repulsion between the lone pair and bonding electrons within each TeO$_4$ unit leads to the corrugated layer structure.
}
\end{figure}
\CTSO~crystallizes in the orthorhombic space group $Pnma$ with two Cu, one Te, one S, and seven O sites.
Figure~\ref{fig:CIF}a shows that each corrugated plane in the quasi-2D structure of \CTSO\ is made of corner-sharing CuO$_4$ squares (Cu1 site) along both $b$ and $c$-directions as well as edge-sharing CuO$_4$ squares and TeO$_4$ pyramids along the $\langle$011$\rangle$ directions.
Figure~\ref{fig:CIF}b shows that the corrugated layers are linked by slabs of interconnected corner-sharing CuO$_4$ (Cu2 site) and SO$_4$ units.
The orientation of these slabs alternates between [10$\mathrm{\bar{1}}$] and [101].
The H$_2$O molecules are inserted between the slabs and diffused along $b$-axis.

Figure~\ref{fig:CIF}c shows the local environment of distorted square-planar CuO$_4$ with bond lengths ranging from 1.9198(13)~\AA\ to 2.0176(14)~\AA, in agreement with the Cu-O bond lengths of 1.9~\AA\ to 2.1~\AA\ in Cu$^{2+}$ compounds Cu$_7$TeO$_4$(SO$_4$)$_5\cdot$KCl~\cite{Pertlik_1988_CTOSOKCl} and Cu$_7$(TeO$_3$)$_2$(SO$_4$)$_2$(OH)$_6$~\cite{Guo_2017_CTOSOOH}.
The O-Cu-O bond angles in the CuO$_4$ units range from 81.15(7)$^\circ$ to 100.5(6)$^\circ$, comparable to the bond angles of 79$^\circ$ to 95$^\circ$ in Cu$_7$TeO$_4$(SO$_4$)$_5\cdot$KCl~\cite{Pertlik_1988_CTOSOKCl}, and 75.0(2)$^\circ$ to 98.3(2)$^\circ$ in Cu$_2$Te$_3$O$_8$~\cite{Christopher_1999_M2Te3O8}.
We present a detailed analysis of the bond valence sum (BVS) using the local coordination in the Supporting Information (Table~S1).
A summary of those results is presented in Table~\ref{tab:tab4}.
The bond valence around the Cu1 site sums to 1.922, confirming a 2+ state.
However, the bond valence around the Cu2 site sums to 1.480 assuming a 1+ state and 1.746 assuming a 2+ state, indicating a mixed valence of Cu$^{2+}$/Cu$^{+}$ for the Cu2 site.

\begin{table}
\caption{\label{tab:tab4}Bond valence sum values for the cation sites in \CTSO.}
\begin{tabular}{ccc}
\hline
\hline
Atomic site & BVS & Expected value\\
  \hline
Cu1 & 1.922 & +2\\
Cu2 & 1.746/1.480 & +2/+1\\
Te & 3.766 & +4\\
S & 6.038 & +6\\
\hline
\hline
\end{tabular}
\end{table}

The Te-O bond distances and the coordination environment for Te is consistent with a Te(IV) oxidation state and comparable to the Te-O bond length range of 1.9 \AA~to 2.1 \AA~in trigonal bipyramidal TeO$_4$ in $\alpha$-TeO$_2$~\cite{LINDQVIST_1968_TeO2}.
The trigonal bipyramid coordination for Te is rarely observed in copper tellurium sulfates and, to the best of our knowledge, was only reported as tetragonal pyramid in Cu$_7$TeO$_4$(SO$_4$)$_5\cdot$KCl~\cite{Pertlik_1988_CTOSOKCl}.
The average Te-O bond lengths of the title compound range from 1.9087(14)~\AA\ to 2.1285(18)~\AA\ comparable to several other copper tellurium oxides such as Cu$_7$TeO$_4$(SO$_4$)$_5\cdot$KCl, Ba$_2$Cu$_2$Te$_2$P$_2$O$_{13}$~\cite{XIA_2015_Ba2Cu2Te2P2O13}, Cu$_2$Te$_3$O$_8$~\cite{Christopher_1999_M2Te3O8}, Nb$_2$Te$_4$O$_{13}$~\cite{Jiang_2004_Nb2Te4O13}, and BaCuTeO$_3$TeO$_4$~\cite{Sedello_1996_BaCuTeO3TeO4}.
%
The O$_\mathrm{ax}$-Te-O$_\mathrm{eq}$ (axial and equatorial oxygens) bond angles in \CTSO\ of 80.23(5)$^\circ$ and 77.29(5)$^\circ$ are comparable to the pairs of tetragonal pyramidal bond angles of 90.5$^\circ$ and 87.9$^\circ$ in Cu$_7$TeO$_4$(SO$_4$)$_5\cdot$KCl~\cite{Pertlik_1988_CTOSOKCl}.
The reduction in the bond angle in the title compound can be due to the repulsion between the lone pair and the bonding electrons in each TeO$_4$ unit within the corrugated layer (Figure~\ref{fig:CIF}c).

\section{\label{thermal}Thermal Analysis}
\begin{figure}
\includegraphics[width=0.55\textwidth]{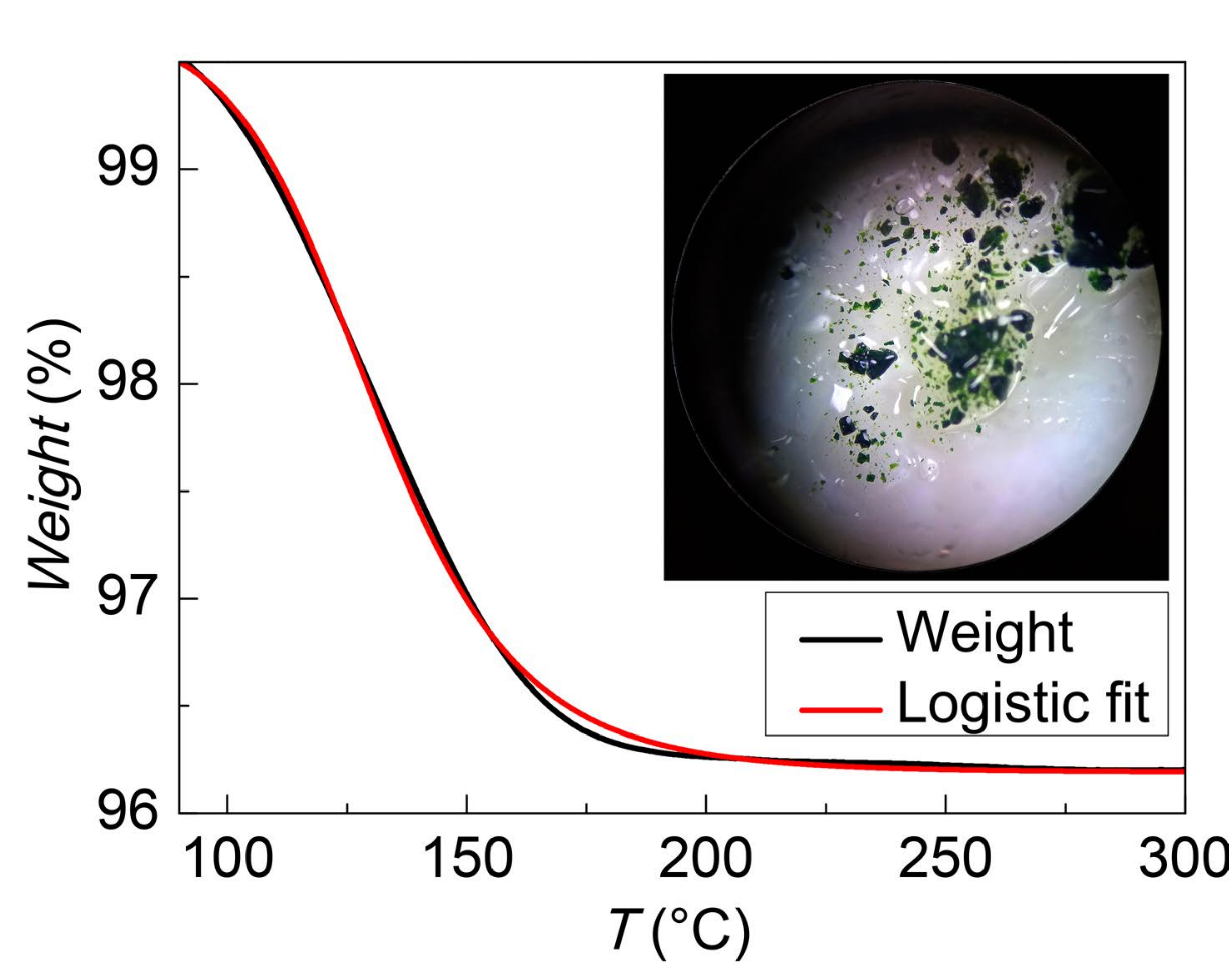}
\caption{\label{fig:TGA}
TGA curve (black) of \CTSO\ and a logistic fit (red) with a good quality ($R^2>$0.998). Inset: crystals after dehydration.
}
\end{figure}

Figure \ref{fig:TGA} shows the change in the mass of \CTSO\ with increasing temperature up to 300\C\ under a nitrogen atmosphere. 
A weight loss of 3.4(1) wt.-\% is observed, which is in agreement with the expected 3.6 wt.-\% corresponding to one H$_2$O molecule per formula unit. 
We fitted the TGA data using a logistic function, $p_w= B+\frac{A-B}{1+ \left(\frac{T}{T_0}\right)^p}$, where $A$ and $B$ are the initial and final weight percentages in the dehydration experiment. 
$T_0$ is the temperature, at which half of the water has been removed and $p$ determines the rate of the dehydration. 
We determined the midpoint of the dehydration to be $T_0$=131(1)~\C\ and the parameter $p$ was 8.620(3). 
The 3.4(1)~\% weight loss was found by subtracting the fit parameter $A$=99.6327(4)\% from $B$=96.1918(1)\%.
Interestingly, the sample turned from black to transparent dark green during the experiment, while the crystal structure stays intact.
This could indicate a change in the mixed oxidations state of Cu2, in favor of Cu$^{2+}$.
The most likely scenario is that Cu$^+$ ions between the layers (Cu2 site) are conjugated with hydronium ions (H$_3$O$^+$) to maintain charge neutrality. Upon dehydration however, most of the Cu$^+$ will turn into Cu$^{2+}$.

\section{\label{magnetic}Magnetic Properties}

\begin{figure}
\includegraphics[width=0.55\textwidth]{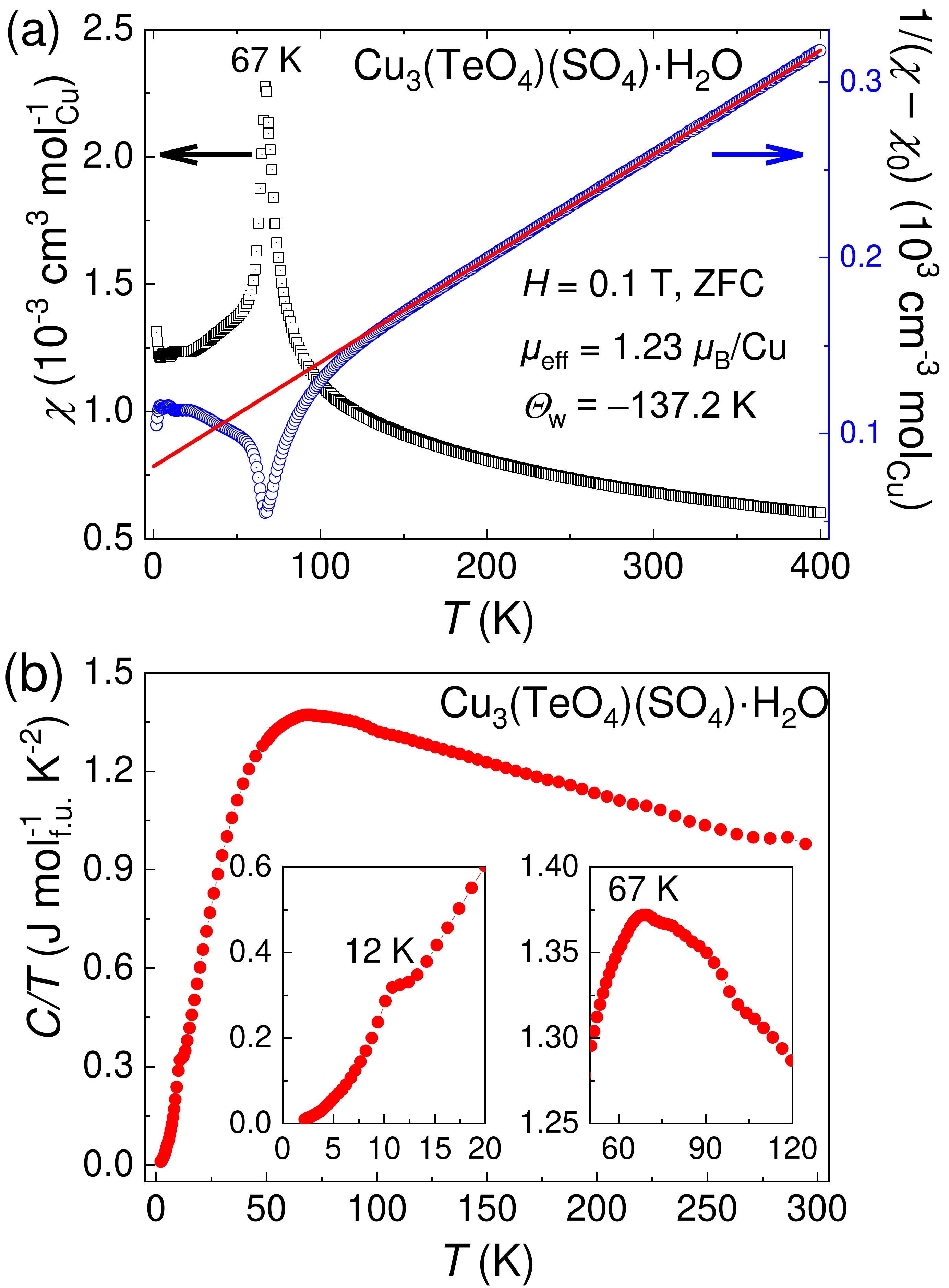}
\caption{\label{fig:CW}
(a) Temperature dependence of the magnetic susceptibility (left axis) and inverse susceptibility (right axis) in \CTSO\ under a magnetic field of 0.1~T in zero-field-cooling (ZFC) mode.
The red solid line is a Curie-Weiss fit to the data between 250 and 400~K.
Due to the acicular habit of crystals, the field direction is not specified in Figures~\ref{fig:CW}, \ref{fig:XT}, and \ref{fig:MH}.
(b) $C/T$ under zero-field as a function of temperature.
The insets magnify the transitions at $T^\star=12$~K (left) and $T_N=67$~K (right).
}
\end{figure}

Figure \ref{fig:CW}a displays the temperature dependence of the magnetic susceptibility $\chi(T)$ in \CTSO\ under an external field of 0.1~T.
A sharp peak at $T_N=67(1)$~K suggests an AFM phase transition.
The inverse susceptibility is fitted to the Curie-Weiss expression $\chi=C/(T-\Theta_\mathrm{W})+\chi_0$, using the data between 250 and 400~K.
From this fit, we obtain a negative Weiss temperature $\Theta_\mathrm{W}$=$-137$~K, confirming AFM correlations above $T_N$.
The effective moment extracted from the Curie-Weiss fit is $1.23~\mu_\mathrm{B}$/Cu, which is 71\% of the expected value ($1.73~\mu_\mathrm{B}$/Cu) for Cu$^{2+}$ (g=2, $S$=1/2).
Based on the BVS analysis mentioned earlier, the mixed valence of the Cu2 site (inter-layer coppers) is responsible for the reduced moment.
Since the occupancy of the Cu1 site ($8d$) is twice that of Cu2 site ($4c$), the reduced moment corresponds to 88\% Cu$^+$ in the Cu2 site.
Thus, most of the inter-layer copper ions are non-magnetic, which makes the analogy between \CTSO\ and superconducting cuprates more justified.
In cuprates, the inter-layer ions are non-magnetic e.g. La$^{3+}$ and Sr$^{2+}$. 
Based on the Curie-Weiss fit analysis and the obtained magnetic moment, the exact formula can be represented as Cu$^{2+}_{3-x}$Cu$^{+}_{x}$(TeO$_4$)$^{4-}$(SO$_4$)$^{2-}$(H$_2$O)$_{1-x}$(H$_3$O$^+$)$_x$, where $x\approx 0.88$. The hydronium ion (H$_3$O$^+$) is introduced to balance the charge, as we mentioned earlier. Nevertheless, we use the formula \CTSO\ for a simple presentation.


The heat capacity data shown in Figure~\ref{fig:CW}b are consistent with the magnetic susceptibility data.
A peak at 67~K in the right inset of Figure~\ref{fig:CW}b is clearly discernible even with the dominant phonon background in the high-temperature regime.
It confirms the AFM transition observed in the susceptibility data at 67~K in Figure~\ref{fig:CW}a.
In addition, a kink is observed in the heat capacity at $T^\star$=12~K, which is magnified in the left inset of Figure~\ref{fig:CW}b.
As explained below, we attribute this feature to a spin-canting transition using the field dependence of magnetization.
%

\begin{figure}
\includegraphics[width=0.55\textwidth]{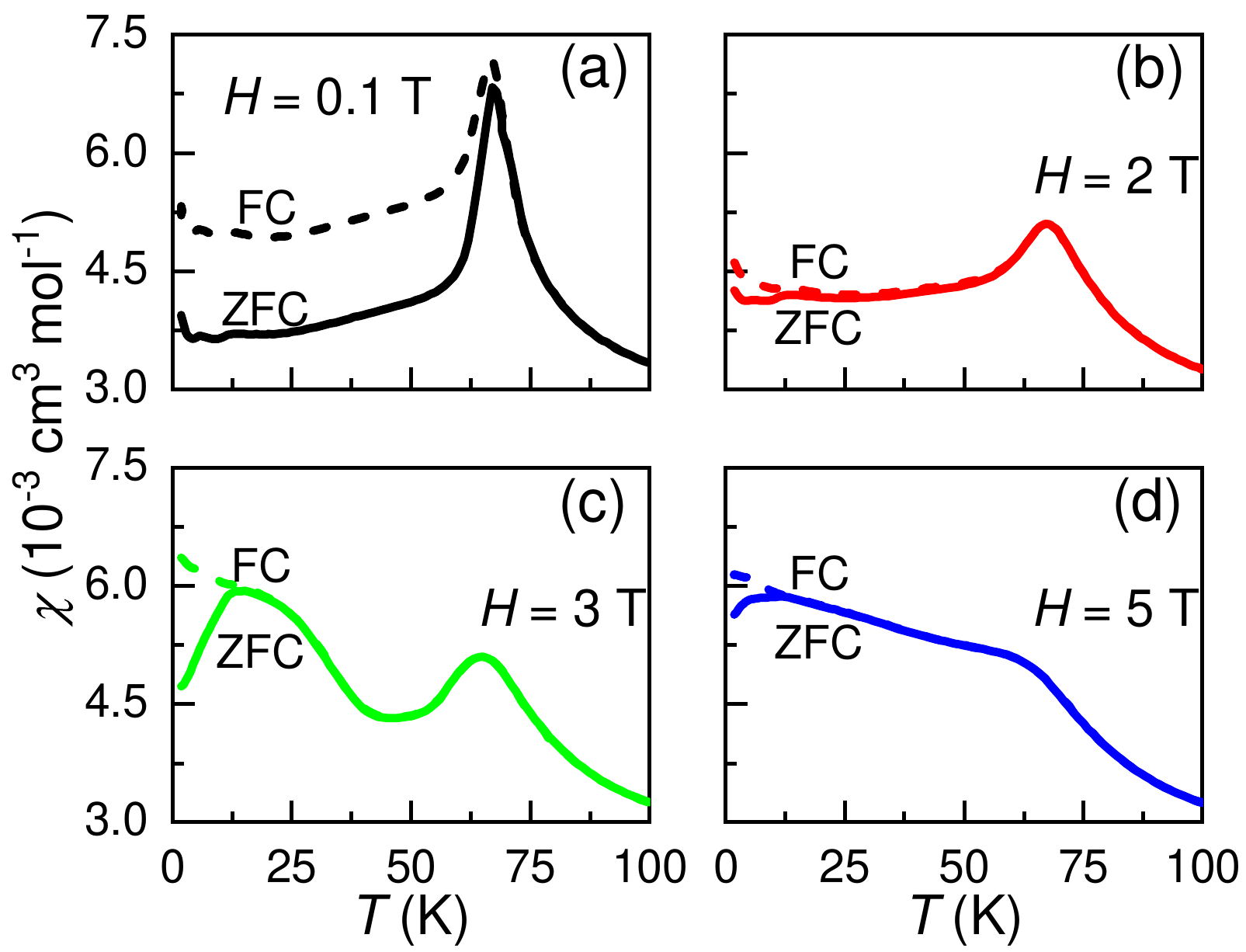}
\caption{\label{fig:XT}
Temperature dependence of the magnetic susceptibility below 100~K under different fields at (a) 0.1~T, (b) 2~T, (c) 3~T, and (d) 5~T.
The solid and dashed lines represent the ZFC and FC conditions, respectively.
}
\end{figure}

To examine the phase transition at $T^\star$=12~K, we studied the field dependence of the magnetic susceptibility under zero-field-cooled (ZFC) and field-cooled (FC) conditions (Figure~\ref{fig:XT}).
Under a small magnetic field (less than 1~T), the AFM peak is accompanied by a splitting between the ZFC and FC data in Figure~\ref{fig:XT}a, suggesting a small ferromagnetic component in addition to the obvious AFM order.
Such a behavior can result from a small deviation from the strictly antiparallel arrangement of spins in a canted AFM~\cite{cao_spin-canted_2017}.
With increasing field to 2~T, the ZFC/FC splitting disappears below $T_N$=67~K, but remains visible below $T^\star$=12~K.
This trend becomes even clearer at 3~T, where two peaks are observed: one at 67~K without the ZFC/FC splitting, and another at 12~K with the splitting.
The splitting below 12~K survives up to 5~T as shown in Figure~\ref{fig:XT}d.
We interpret this behavior as a N\'{e}el-type transition at $T_N$=67~K followed by a spin-canting transition below $T^\star$=12~K.


\begin{figure*}
\includegraphics[width=0.9\textwidth]{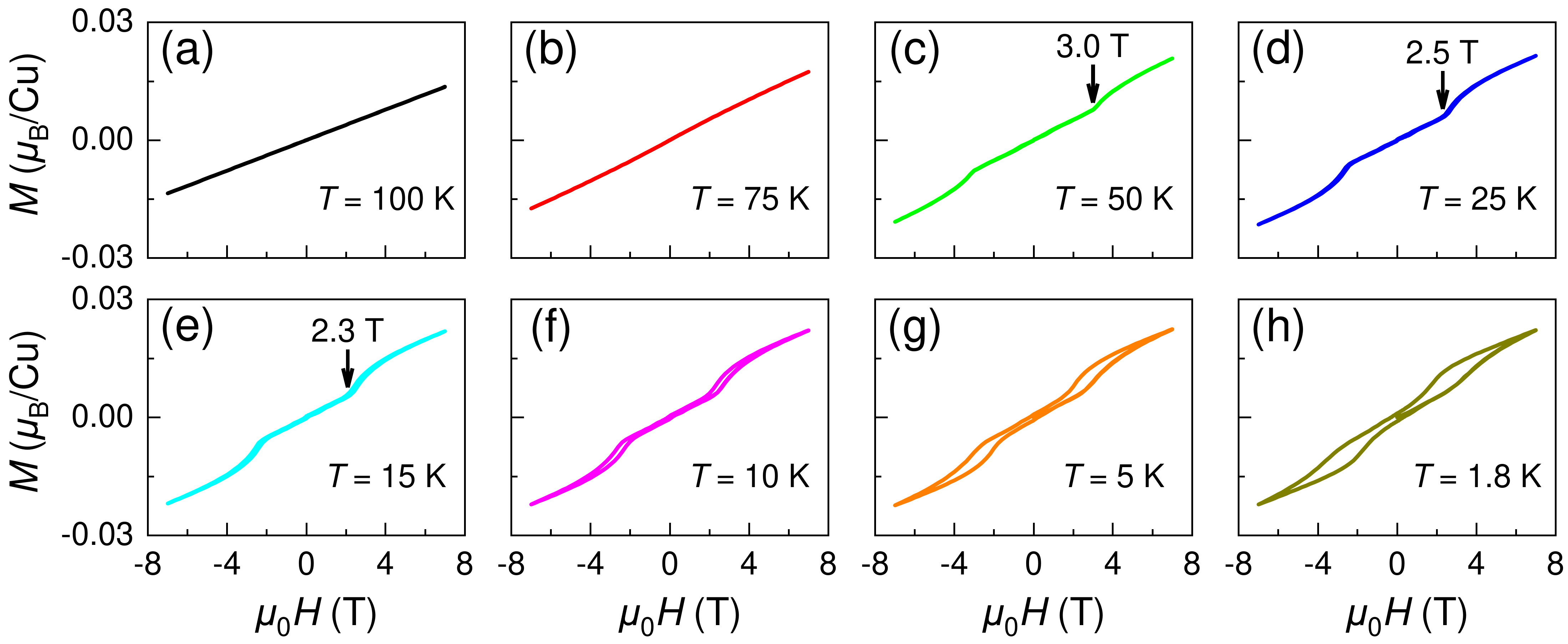}
\caption{\label{fig:MH}
Isothermal magnetization loops in \CTSO~at several temperatures (a,b) above $T_N$, (c,d,e) between $T^\star$=12~K and $T_N$=67~K, and (f,g,h) below $T^\star$.
}
\end{figure*}

Further evidence of a spin-canting transition at 12~K comes from the evolution of isothermal magnetization loops in Figure~\ref{fig:MH}.
At $T$$>$$T_N$, the $M(H)$ curves are linear as seen in Figure~\ref{fig:MH}a,b.
With decreasing temperature below $T_N$=67~K, a small step-like increase is observed in the $M(H)$ curves at a critical magnetic field which is moderately suppressed from $H_c$=3 to 2.3~T as the temperature is decreased from 50 to 15~K.
The step-like transition could be a mild spin-flop in the AFM ordered Cu$^{2+}$ moments.
A similar phenomenon is reported in Ni$_3$TeO$_6$\cite{Oh_2014_NTO}, another layered material containing three Ni sites.
With further decreasing temperature below $T^\star$=12~K, a hysteresis loop opens in the $M(H)$ curves, confirming a FM component in the magnetic order as expected from a spin-canting transition~\cite{li_spin_2009}.
Note that the spin-canting transition and the resulting FM component is well justified by the presence of two competing coupling constants $J_1$ and $J_2$ (Figure~\ref{fig:MOTIVATION}b).

\section{Conclusions}
The salient features of \CTSO\ can be summarized as follows.
It is a layered material with spin-1/2 Cu$^{2+}$ ions within the layers and mixed valence Cu$^{2+}$/Cu$^+$ between the layers.
The layers are made of corner-sharing CuO$_4$ square-planar units in the $bc$ plane, which are corrugated due to edge-sharing between the CuO$_4$ and TeO$_4$ units.
These features are reminiscent of the high-$T_c$ cuprate superconductors~\cite{guo_electronic_1988}, but unlike the cuprates, the layers are corrugated in \CTSO.
Despite tremendous effort by materials experts, it has been difficult to identify new families of compounds with such structural motifs and a Mott insulating AFM ground state.
In this regards, \CTSO\ is a promising candidate material which is also available in single crystal form.
The AFM transition temperature of 67~K and the half-filling of Cu$^{2+}$ confirm a Mott insulating ground state.
The spin-canting transition at 12~K requires at least two different super-exchange couplings $J_1$ and $J_2$, consistent with the different bond lengths in Figure~\ref{fig:MOTIVATION}b.
A $J_1$/$J_2$ magnetic model has been theoretically proposed to harbor both superconductivity and spin liquid behavior in cuprates~\cite{yu_spin-frac12_2012,yu_deconfinement_2018}.
However, the largest $J_1/J_2$ ratio in cuprates is around 0.5, since $J_2$ is a next-nearest-neighbor interaction~\cite{yu_spin-frac12_2012,yu_deconfinement_2018,tanaka_effects_2004}.
Remarkably, \CTSO\ provides access to a $J_1/J_2$ ratio close to 1, due to the small difference between the $J_1$ and $J_2$ super-exchange paths (Figure~\ref{fig:MOTIVATION}b).
Evidence of a mild magnetic frustration, due to the competition between $J_1$ and $J_2$, can be observed in Figure~\ref{fig:CW}a, where the Weiss temperature ($\Theta_W$=$-137$~K) is twice as large as the N\'{e}el temperature ($T_N$=67~K).
Future research directions from here would be to find whether $T_N$ can be suppressed under pressure, leading to a spin liquid ground state, or if it can be suppressed by doping (e.g. replacing the interlayer Cu atoms with Ag, Zn, and Ca) to induce superconductivity.
It will also be instructive to synthesize a deuterated version of \CTSO\ to solve the magnetic structure using neutron diffraction.
The intricate chemistry of tellurite-sulfate systems and the versatility of the hydrothermal method are likely to produce more such materials in the future.

\begin{acknowledgement}
The work at Boston College was funded by the National Science Foundation under award number DMR--1708929.
The work at UT Dallas was funded by the National Science Foundation under award number DMR--1700030.
We acknowledge the support of the National Institute of Standards and Technology in providing the neutron research facilities used in this work. The identification of any commercial product or trade name does not imply endorsement or recommendation by the National Institute of Standards and Technology.
\end{acknowledgement}

\begin{suppinfo}


\begin{itemize}
  \item Wang-15apr2021-SI.pdf: supporting information.
  \item X-ray crystallographic information file (CIF) for \CTSO\ is available at \url{http://www.ccdc.cam.ac.uk/data_request/cif}. The associated deposition number is CSD 2076676.
\end{itemize}

\end{suppinfo}

\bibliography{Wang_8jun2021}

\pagebreak
 \begin{figure}
 \includegraphics[width=0.55\textwidth]{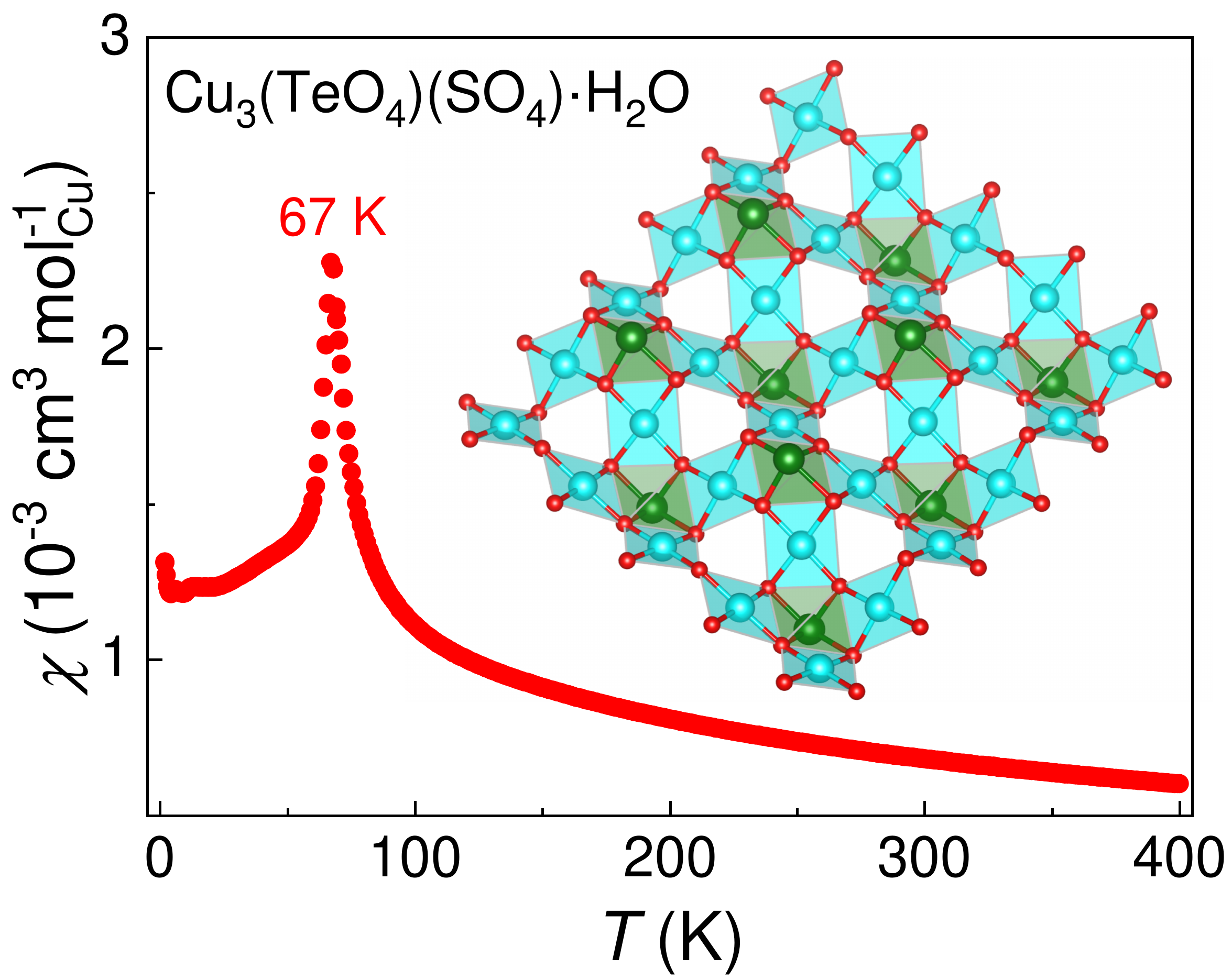}
 \caption{\label{TOC}
 A new antiferromagnetic material is made with a corrugated layered crystal structure and transition temperature of 67 K.  
 The corrugated layer comprises corner-sharing CuO$_4$ squares and edge-sharing CuO$_4$-TeO$_4$ units. 
 The structural similarity between \CTSO\ and the cuprate materials makes \CTSO\ a promising platform for finding superconductivity, spin liquid, or other exotic quantum states.}
 \end{figure}

\end{document}







\pagebreak
\section{Energy Dispersive X-ray (EDX) Spectroscopy}
\begin{figure}[!htb]
 \includegraphics[width=0.6\textwidth]{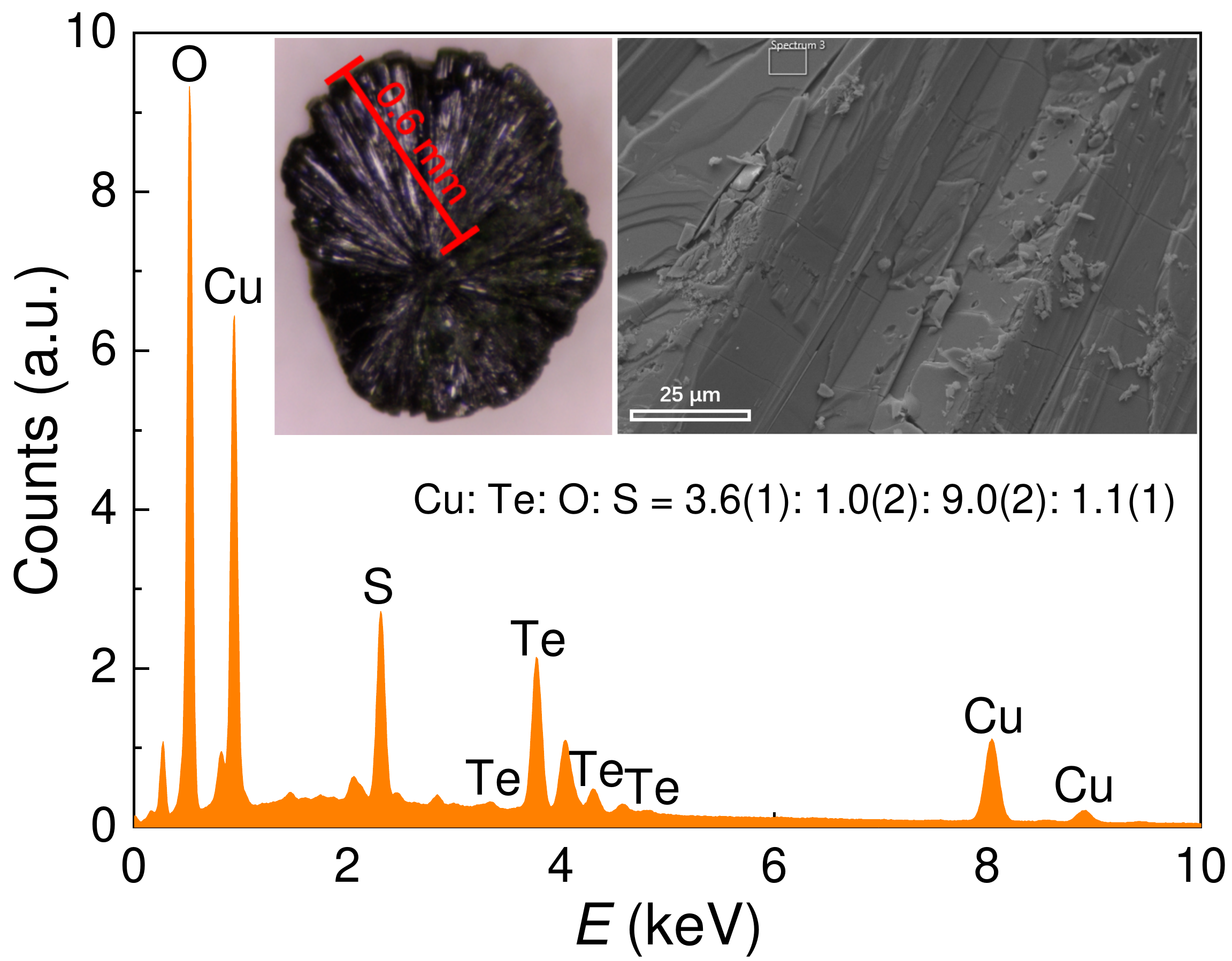}
 \caption{\label{fig:EDX}
The energy dispersive X-ray spectroscopy with the electron beam focus on the rectangular spot of the crystal's fresh surface (see the inset). The atom ratio averaged with several spots is also shown in the figure.
 }
 \end{figure}

The results of EDX spectroscopy are presented in Figure~\ref{fig:EDX}, confirming the stoicheometry of the title compound. 

\section{Neutron Diffraction}

We presented the details of single crystal X-ray diffraction in the main text (Tables 1 and 2).
We also performed neutron scattering experiments, primarily to determine the magnetic structure.
Unfortunately, the magnetic Bragg peaks could not be identified reliably due to a large background from the incoherent scattering of neutrons by hydrogen atoms in \CTSO.
However, the neutron diffraction patterns helped confirming the structural refinement from X-ray diffraction and ruled out structural transitions at low temperatures.

\begin{figure}[!htb]
 \includegraphics[width=0.65\textwidth]{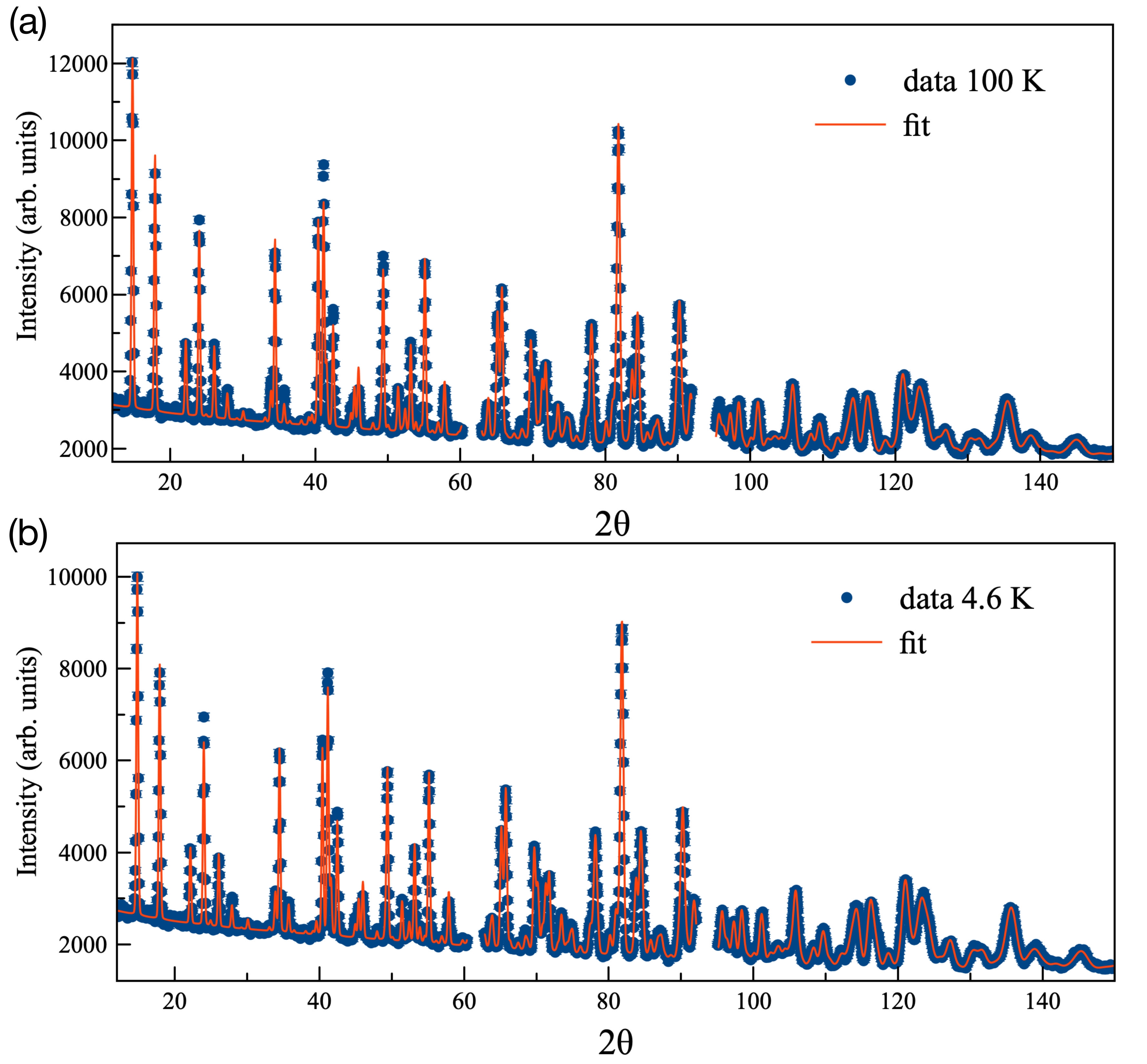}
 \caption{\label{fig:N}
Neutron diffraction pattern on a powder specimen of \CTSO\ at (a) 100~K and (b) 4~K. Data points are presented in blue and the Rietveld fit is in red. 
 }
 \end{figure}

The measured neutron diffraction patterns are shown in Figure~\ref{fig:N}a and b for $T$=100 and 4.6~K, respectively. 
Using the 100~K data, we first confirmed the nuclear structural parameters of \CTSO\ and confirmed our single-crystal X-ray refinement. 
To do so, we performed the neutron diffraction refinement using GSAS~\cite{toby_gsas-ii_2013} by assuming the structural parameters from the X-ray refinement (Tables 1 and 2 in the main text). 
We obtained the refinement in Figure~\ref{fig:N}a with $\chi^2$=3.1 by solely allowing the scale factor, instrument resolution parameters, lattice parameters, and the background parameters to vary. 
We did not detect magnetic Bragg peaks at 4.6~K due to a combination of weak magnetic Bragg peaks lying on top of a high incoherent background from the hydrogen atoms. 
Importantly, we did not detect any splitting of the structural Bragg peaks within a Q-resolution of ~0.015 Å-1 so that no structural transition was observed at low temperature. 
Indeed, assuming slightly different lattice parameters, the 4.6~K diffraction pattern (Figure~\ref{fig:N}b) can be reproduced using the same structural parameters as the 100~K fit.

\section{Bond Valence Sum Calculations}
The bond valence sum (BVS) values were calculated using the implementation in WinGX by Brown and Altermatt~\cite{Brown_BVS_1985} based on the provided reference bond distances $r_o$~\cite{Brese_BVS_1991}.
At a first glance, the BVS values simply indicate the oxidation states Cu(II) for both Cu-sites, Te (IV), and S(VI). 
However, the BVS calculations around Cu1 include two bonds to Te (ca. 3~\AA) and a fifth O-bond to O2 for Cu2, which is much longer than the other Cu-O bonds (ca. 2.4~\AA~compared to ca. 2.0~\AA, respectively). 
If we only focus on the distorted square planar coordination of Cu, the BVS values differ significantly despite similar coordination environments. 
The BVSs are 1.922 and 1.746 and for Cu1 and Cu2, respectively. 
The reduced BVS values of Cu2 could indicate a mixed valence of Cu(I)/Cu(II) on this site as inferred from the measured magnetic moment. 
Table \ref{tab6} shows the calculated BVS values for the atomic sites compared to their expected values. 
We do not find any indications for oxygen vacancies in the signle crystal X-ray data (confirmed by neutron diffraction) that could balance the reduced charge on Cu2. 
Therefore, we attribute the reduced charge on Cu2 and the mixed Cu$^+$/Cu$^{2+}$ valence to the presence of hydronium ion between the layers, i.e. a mixture of H$_3$O$^+$/H$_2$O.
The BVS value of Te also deviates from the expected value, which could indicate that the presence of Cu(I) is compensated by Te(VI).

\begin{table}
\caption{\label{tab6}Bond Valence Sum for the compound \CTSO.}
\begin{tabular}{ccccc}
\hline
\hline
Atom & Observed Distance (\AA) & $r_0$ (\AA) & B.val. & Sum\\
  \hline
\multicolumn{5}{l}{Te: assuming a valence of 4+ for Te}\\
O1 & 2.1287 & 1.955 & 0.674 & 0.674\\
O2 & 2.0165 & 1.955 & 0.870 & 1.543\\
O3 & 1.9085 & 1.955 & 1.112 & 2.655\\
O3 & 1.9085 & 1.955 & 1.112 & 3.766\\
\multicolumn{5}{l}{Bond valence sum = 3.766; discrepancy = 0.234}\\
\\
\multicolumn{5}{l}{Cu1: assuming a valence of 2+ for Cu1}\\
O1 & 1.9628 & 1.679 & 0.455 & 0.455\\
O2 & 1.9534 & 1.679 & 0.467 & 0.922\\
O3 & 1.9372 & 1.679 & 0.488 & 1.410\\
O3 & 1.9198 & 1.679 & 0.512 & 1.922\\
\multicolumn{5}{l}{Bond valence sum = 1.922; discrepancy = 0.078}\\
\\
\multicolumn{5}{l}{Cu2: assuming a valence of 1+ for Cu2}\\
O1 & 1.9283 & 1.610 & 0.423 & 0.423\\
O4 & 2.0174 & 1.610 & 0.332 & 0.755\\
O4 & 2.0174 & 1.610 & 0.332 & 1.087\\
O5 & 1.9556 & 1.610 & 0.393 & 1.480\\
\multicolumn{5}{l}{Bond valence sum = 1.480; discrepancy = 0.48}\\
\\
\multicolumn{5}{l}{Cu2: assuming a valence of 2+ for Cu2}\\
O1 & 1.9283 & 1.679 & 0.500 & 0.500\\
O4 & 2.0174 & 1.679 & 0.391 & 0.891\\
O4 & 2.0174 & 1.679 & 0.391 & 1.282\\
O5 & 1.9556 & 1.679 & 0.464 & 1.746\\
\multicolumn{5}{l}{Bond valence sum = 1.746; discrepancy = 0.254}\\
\\
\multicolumn{5}{l}{S: assuming a valence of 6+ for S}\\
O4 & 1.4802 & 1.624 & 1.475 & 1.475\\
O4 & 1.4802 & 1.624 & 1.475 & 2.950\\
O5 & 1.4714 & 1.624 & 1.511 & 4.461\\
O6 & 1.4553 & 1.624 & 1.577 & 6.038\\
\multicolumn{5}{l}{Bond valence sum = 6.038; discrepancy = 0.038}\\
\hline
\hline
\end{tabular}
\end{table}

\pagebreak
\bibliography{Wang_8jun2021_SI}
